\documentclass[conference]{IEEEtran}
\IEEEoverridecommandlockouts
\usepackage{cite}
\usepackage{amsmath,amssymb,amsfonts}
\usepackage{algorithmic}
\usepackage{graphicx}
\usepackage{textcomp}
\usepackage{xcolor}
\def\BibTeX{{\rm B\kern-.05em{\sc i\kern-.025em b}\kern-.08em
    T\kern-.1667em\lower.7ex\hbox{E}\kern-.125emX}}

\usepackage{graphicx}
\usepackage[linesnumbered,ruled,vlined]{algorithm2e}

\begin{document}

\title{Social Recommendation through Heterogeneous Graph Modeling of the Long-term and Short-term Preference Defined by Dynamic Time Spans \\
}

\author{\IEEEauthorblockN{Behafarid Mohammad Jafari, Xiao Luo, Ali Jafari}
\IEEEauthorblockA{\textit{Purdue School of Engineering and Technology} \\
\textit{Indiana University - Purdue University Indianapolis}\\
Indianapolis, USA \\
bmohamma@purdue.edu, luo25@iupui.edu, jafari@iupui.edu}

}

\maketitle

\begin{abstract}
Social recommendations have been widely adopted in substantial domains. Recently, graph
neural networks (GNN) have been employed in recommender systems due to their success in graph representation
learning. However, dealing with the dynamic property of social network data is a challenge. This research presents a novel method that provides social recommendations by incorporating the dynamic property of social network data in a
heterogeneous graph. The model aims to capture user preference over time without going through the complexities of a dynamic graph by adding period nodes to define users’ long-term and short-term preferences and aggregating assigned edge weights. The model is applied to real-world data to argue its superior performance. Promising results demonstrate the effectiveness of this model\footnote{Source Code: https://github.com/BehafaridMjf/Social-Recommendation-System}.
\end{abstract}


\section{Introduction}
A substantial number of recommendation algorithms have been developed and evaluated recently \cite{keyw}. Content-based filtering,  collaborative filtering, and hybrid methods are usually considered as the traditional methods \cite{three}. Recent advances in deep learning-based recommender systems have gained significant attention due to addressing issues in conventional models and achieving high recommendation quality \cite{advdeep}. However, relatively little attention has been given to modeling the dynamic behavior of the users in social recommender systems. Introducing the dynamic property of social networks to graphs can help improve any platform that uses a social recommender system including CourseNetworking\cite{rumi} as a social learning management system.

Recommendation algorithms are built to learn the preferences and interests of the users \cite{cn}. However, using historical data to predict future behavior is the cornerstone of many recommendation systems. Past user behavior sequences can reflect the user’s future interaction \cite{historic}. The tastes of users may change over time and recommender systems should capture these dynamics and stay tuned to the latest user interests\cite{detect}. However, most existing models attempt to capture preferences using a static view \cite{staticview} due to the complexities of dynamic models. 

Most traditional and deep learning-based recommendation systems do not account for short-term and long-term interests at the same time. Long-term interest may be successful in determining the overall behavior of a user. But at any given time, user preference is also affected by short-term interest \cite{stg}. In this paper, we propose a method to consider both the short-term and long-term interest in one graph by adding time span nodes to model the dynamics of the user preferences over time.

Social networks are represented by graph structures \cite{socialgraphs}. With graphs becoming richer with information and also due to their unstructured nature, most machine learning algorithms are not applicable or efficient when working with the graphs \cite{amir}. To deal with this challenge, Graph Neural Network(GNN) \cite{gnn} is employed in our proposed model which is proven to be highly efficient on social graphs.

In this paper, we propose a novel approach to deal with the challenge of user's dynamic behavior and capture short-term and long-term preferences using the GNN algorithm.
Our model has three main superiorities over the existing methods:

\begin{itemize}
    \item Our proposed model considers the evolution of user interest associated with a defined time span. It considers short-term preferences in addition to long-term interest and aggregates both when predicting the user's future behavior. The time span can be defined according to the application's needs. 
    
    \item Our approach utilizes information from all possible edges and nodes in the graph including users, items, and time spans. So that the model can generate recommendations by considering all entities and edges.
    
    \item Our model builds with the attention mechanism to incorporate the weighted edges between users, items, and time spans for prediction and shows good performance on multiple data sets.
    
    \item The extensive model analysis shows the contributions of different components of our model. This will shade lights for future research in building complex graph models to consider the time dimension.   

\end{itemize}

\section{Related Works}
The mainstream modeling paradigm in recommender systems has developed
from neighborhood methods to representation learning-based frameworks \cite{neighbtograph}. The neighborhood methods make recommendations by directly using historical data and interactions.
However, representation learning-based methods try to encode entities as
continuous vectors in a shared space (i.e., embeddings). DeepWalk \cite{deepwalk} and Node2Vec \cite{node2vec} are examples of algorithms that embed nodes into vectors based on the probability of node co-occurrences in a random walk \cite{randomwalk}. There are several different methods proposed to learn the representations of the entities, such as matrix factorization \cite{mf}, and deep learning models including graph learning-based methods \cite{glb}, which consider the information from the perspective of graphs \cite{neighbtograph}.

Deep learning often outperforms traditional methods due to the capability to process non-linear data and nontrivial relationships between entities. The most relevant work with neural networks includes DeepSoR \cite{deepsor}, which integrates
neural networks into probabilistic matrix factorization. Recently, graph learning-based methods \cite{glb} have made great improvements in machine learning.
Graph Neural Networks (GNN) \cite{gnn} was proposed to directly process graphs. GNN extends recursive neural networks and can deal with most kinds of graphs. GCMC \cite{gcmc} is another model that uses the convolutional layers. ScAN \cite{scan} employs the co-attention mechanism for making recommendations. GraphRec \cite{graphrec} also benefits from the attention mechanism. It uses the attention network to collect information about the entities which have mutual interactions in addition to what ScAN does. To improve the performance of the recommendation system, aggregating information from all possible sources when building connections is an essential step. HeteroGraphRec \cite{amir} uses modern neural network structures in a social network model to aggregate both the item-related and user-related information in a social network. This model considers items in a social network as nodes in the graph and it contains item–item connections created based on the similarity and then applies the attention mechanism to make recommendations. CapsRec \cite{capsrec} is another method based on the combination of capsule network and GNN for social recommendations, which considers biases in enriching user/item latent factor vectors. Many of these GNN-based models do not consider the user preference over time in the graph.

In most real-world social networks, the users' preferences are changing over time. 
A time-varying social network is proposed \cite{timebin} to consider temporal changes over time. The model divides the data into bins for each time window, creates a graph for each window, and then connects the neighboring graphs using forward and backward edges. It finally uses randoms walk \cite{randomwalk} for making recommendations. A Session-based Temporal Graph (STG) \cite{stg} was created to address the problems that existed in the previous method by considering long-term and short-term preferences benefiting from the time bins and using one static graph instead of multiple graphs. The model employs Injected Preference Fusion (IPF) and extended personalized random walk for temporal recommendation. Although both models consider time bins to improve performance, none of them benefits from high-performance deep learning models when making the recommendations. Long-term and short-term preferences were also used in LSTPM model \cite{poi} which captures long-term preference with a nonlocal network and short-term interest using a geo-dilated RNN or ILSTP model \cite{SR} which combines time interval embedding function,
self-attention, hybrid aggregator, gated graph neural network,
and attention aggregator to integrate long-term and short-term preferences. However, the LSTPM model is developed for location-based networks and recommends the next point-of-interest. Also, although ILSTP is a session-based recommender system, it does not consider users' explicit preferences (e.g., user-item ratings).

Different from these existing models, we propose a new model which benefits from GNN and attention mechanism and applies them to a graph with connections between items and time spans to incorporate temporal data to consider users' long-term and short-term preferences. It is worth noting that the length of the time span can be defined dynamically according to the system's needs and domain knowledge.

\section{Proposed Heterogeneous Graph Model for Recommendation}\label{framework}


\subsection{Graph Architecture} \label{social_model}

Static social networks are represented by the social graph $G = (V, E)$, where $V$ denotes the set of vertices or nodes, and $E$ is the set of edges or links representing interactions between entities within a time span. However, users' preferences and interests may vary during a particular period. Hence, the change in short-term or long-term preferences over time cannot be ignored in a social network. For instance, a user who is always interested in action movies decides to try romantic movies for a short period. In this case, action movies remain as the long-term interest of the user and romantic movies would be considered as the short-term interest of the user during that particular period. An efficient social recommendation system should be able to capture long-term and short-term preferences and make predictions considering both.

\begin{figure}[h!]
\centering
    \includegraphics[scale= 0.37]{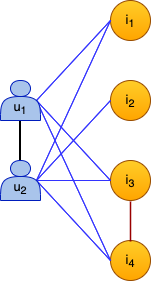}
    \caption{A typical social graph.}
    \label{fig:basic_graph}
\end{figure}
\begin{figure}
    \centering
    \includegraphics[scale = .37]{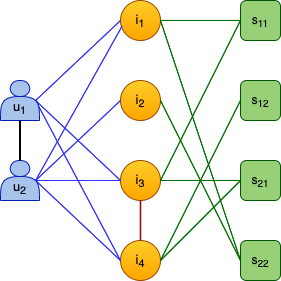}
    \caption{The proposed social graph.}
     \label{fig:proposed_graph}
\end{figure}

The basic and traditional model of the heterogeneous graph is created with two types of nodes shown in Figure \ref{fig:basic_graph}, including users and items and the connections between them including user-user connections representing the trust or social relation between pair of users, user-item edges representing the items each user interacted with, and item-item links connecting similar items. A major limitation is that it does not consider the change in the preferences of the users over each time unit because of the static structure and ignoring time in the graph. 

In this research, we utilize Graph Neural Networks to make predictions by proposing a novel graph architecture to attain rich information by adding the time span nodes. There are three types of nodes in the structure of the proposed heterogeneous graph including users, items, and time span nodes shown in Figure \ref{fig:proposed_graph}.  The relationship between each pair of nodes (if it exists) forms the edges of the graph. The orange circles represent items and the green squares illustrate time span nodes. The black edge(s) introduces the social relations 
between users. The red edges represent the similarities between items. The blue edges define the long-term preferences and the green edges capture the short-term interests during each time span. The time span nodes and the edges between item and time span nodes are designed to capture user preferences over time.

\subsubsection{User-User connection}
 The type of interaction between the users varies for different networks. Thus, there are different indicators to determine the existence of the edges between user nodes. For example,  two users are connected if they share information with each other or if they are set to be friends. 
\subsubsection{User-Item connection}
Users interact with one to many items during the observation interval or defined time span. The user-item edge determines whether a user has interacted with an item.
\subsubsection{Item-Item connection}
The item-item connections are formed when items share particular similar features. This similarity metric can vary from one dataset to another. For example, the items in the same group can be connected to each other. 

\subsubsection{Item-Time span connection}
Each user interacts with an item at a particular time. Based on the timing, the interaction falls in a time span. Each time span can be dynamically defined as a time interval, e.g. one hour or one week, etc.  
The item-time span connection adds time information to the interactions and includes it on the graph. 

As shown in Figure \ref{fig:proposed_graph},
in the proposed graph, the user node connects to all items interacted with by that user, which represents the user’s long-term preferences. Whereas the time span node $ij$ only connects to items which user $i$ interacted with in time span $j$, which represents the user’s short-term preferences during the $j^{th}$ period of time. In Figure \ref{fig:proposed_graph}, there are user $u_1$ and user $u_2$, and items $i_1$, $i_2$, $i_3$, and $i_4$. If the observation time interval is set to be two weeks and the length of a time span is defined as one week, there are two time spans for each user during the observation time: week one ($s_{11}$) and week two ($s_{12}$). User $u_1$ interacts with items $i_1$ and $i_3$ during the time span $s_{11}$, and interacts with items $i_4$ during the $s_{12}$. Whereas user $u_2$ interacts with items $i_3$ and $i_4$ during the week one and $i_1$ and $i_2$ during the week two. The edge between users $u_1$ and $u_2$ shows that they are friends or in each other's trust network, and the edge between items $i_3$ and $i_4$ shows that they are similar items.

Figure \ref{fig:proposed_graph} shows that items $i_1$ and $i_3$ are the short-term interests of user $u_1$ during time span $s_{11}$ and item $i_4$ is the short-term interest during time span $s_{12}$. It must be noted that the time spans are defined to be owned by the user so that they can be dynamically defined according to the user's behaviors. The first argument of the time span $s$ is the corresponding user id who owns the time span, and the second argument of $s$ represents the time span id. For user $u_2$, items $i_3$ and $i_4$ are considered as short-term preference during the time span $s_{21}$, and items $i_1$ and $i_2$ as short-term preference during time span $s_{22}$. However, for long-term interest, we consider all items connected to the same user to determine the long-term preference. Items $i_1$, $i_3$, and $i_4$ are considered as long-term interest of user $u_1$ since user $u_1$ has interactions with all these items during a defined observation interval, though in different time spans. Similarly, items $i_1$, $i_2$, $i_3$, and $i_4$ are considered as long-term interest of user $u_2$.

\subsection{Time Span Construction}
\label{session_construc}

This section explains how time span nodes are formed by dividing an observation time into time spans. To slice the observation time $t$ into time spans, $t$ can be divided into a number ($\frac{t}{pt}$) of equal time spans for each user, where $pt$ can be defined as an hour, a day, a week, a month, or other duration. Time span $s_{ij}$ corresponds to activities of user $i$ during time span $j$. We can start the observation time from a specific timestamp $T$, e.g. when the application launches or a specific date and time. 

Each user $u$ can interact with one or more items -- a item set $I_{us}$ during a time span $s$. For example, item set $I_{ua} =\{i^a_1, i^a_2, ... ,i^a_m \}$ defines the items user $u$ interacted with during the time span $a$, whereas the item set $I_{ub} =\{i^b_1, i^b_2, ... ,i^b_n \}$ includes the items user $u$ interacted with during the time span $b$. Hence, time span $s_{ua}$ has connections to all items in $I_{ua}$, and time span $s_{ub}$ has connections to all items in $I_{ub}$. For the observation time, user $u$ is connected to all elements in both item sets $I_{ua} \cup I_{ub}$.

\subsection{Edge Weights}\label{weights}

To consider the change in user interests over time, edge weights are assigned to the graph. As time goes on, the importance of the items a user interacted with reduces over time. To put in other words, the interactions occurring closer to the current time must get more attention. This information is encoded into the graph using edge weights.

We calculate the time difference between the beginning of the observation $T$ (e.g. starting timestamp of an application) and the interaction timestamp $ts$ to assign weights to the edges. Each interaction between a user $u$ and an item $i$ occurring at timestamp $ts$ defines an edge weight between $u$ and $i$ at $ts$ as $W_{ui}=ts - T$. The edge weight $W_{ui}$ in this study is calculated as the number of seconds. The edge weight $W_{ui}$ increases if there is a new interaction between $u$ and $i$ at a later timestamp since the $ts$ increases over time while $T$ stays the same for all interactions. The hypothesis for this design is that the newer interaction is relatively more important than the older one. As time goes on and new interactions occur, their impact on determining user preference increases. In other words, edge weight increases over time by applying such a calculation. 

The same method is applied while assigning edge weights between item nodes and time span nodes within each time span. Each interaction between a user $u$ and an item $i$ occurs at a timestamp belonging to a time span $s$. The edge weight between $i$ and $s$ at $ts$ is defined as $W_{is}=ts - T$. The edge weight $W_{is}$ in this study is also calculated as the number of seconds.

We then aggregate (shown as Equation \ref{sessionaggre}) the edge weights $W_{ui}$ if the category of the items are similar (e.g., the items are in the same category) or similar based on the definition of each data set. We also aggregate the edge weights $W_{is}$ within the same time spans when the categories of the items are similar, shown as Equation \ref{sessionaggre1}. The objective is to emphasize the relative importance of the interactions among each other based on the timestamp while intensifying the effect of interacting with similar items in determining preferences. The weights aggregation does not cross time spans while generating weights for items to time span edges so that each time span captures the preferences corresponding to that particular period.

\begin{equation}
\label{sessionaggre}
  W_{ui} = \sum_{i_c \in I_u} W_{ui_c}\ \ | \ \ c = category\ \ of\ \ i,\ \ ts_{ui_c} \le ts_{ui}
\end{equation}

\begin{equation}
\label{sessionaggre1}
  W_{is} = \sum_{i_c \in I_s} W_{i_c s}\ \ | \ \ c = category\ \ of\ \ i,\ \ ts_{i_cs} \le ts_{is}
\end{equation}

\begin{figure}
    \centering

    \includegraphics[scale =1.3]{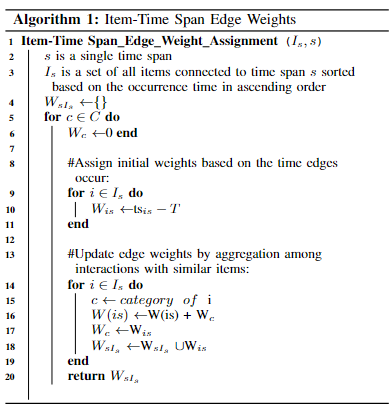}
    \label{fig:enter-label}
\end{figure}

The edge weight assignment between $s$ and $I_s$ (A set of items that are interacted with by user in the time span $s$) is calculated using algorithm \ref{algorithm_weight_assignment1}.   
In algorithm \ref{algorithm_weight_assignment1}, $W_c$ holds the cumulative sum of the weights of the items of category $c$ viewed by the same user $u$ during a specific time span $s$. The $C$ stands for the set of all categories. $W_{sI_s}$ is the set of edge weights for links between time span $s$, and the items interacted by user during time span $s$ known as $I_{s}$. $ts_{is}$ is the timestamp indicating the user's interaction with item $i$ during the time span $s$.

The same algorithm is used to calculate $W_{uI_u}$ for edges between user and item. For each item category that the user previously interacted with, the edge weight between the item and the user is aggregated with the previous edge between the same user and the similar item.

The edge weights in $W_{uI_u}$ and $W_{sIs}$ are normalized so that all fall between 0 and 1 using Equation \ref{norm_weights} and \ref{norm_weights1} respectively.  
In Equation \ref{norm_weights}, $W_{ui}$  and $W_{is}$ are defined as each edge weight in set $W_{uI_u}$ and $W_{sI_s}$ respectively. Also $W_{max}$ is the maximum and $W_{min}$ is the minimum value of the corresponding set for each equation.

  \begin{equation}
  \noindent
    \label{norm_weights}
    (W_{e})_{ui}= \frac{W_{ui} - W_{min}}{W_{max}-W_{min}}
  \end{equation}
  \noindent
  \begin{equation}
    \label{norm_weights1}
    (W_{e})_{is}= \frac{W_{is} - W_{min}}{W_{max}-W_{min}}
  \end{equation}

Our model is designed in a way that different edge weights can be assigned to the user-user and item-item connections. So, the user-user and item-item edge weights can be set to different values to consider more dynamics. In this research, edge weight of the user-user and item-item connections are set to be 1 which means the edge weight parameter in the attention mechanism for these two types of connections is important and considered the same.

\subsection{Attention Mechanism with Edge Weights} \label{aggregation}

After graph construction and edge weights calculation, the constructed graph is trained to generate user, item, and time span embeddings which are then fed to a neural network for prediction.

Specifically, a graph attention network \cite{gatcite} is applied to our heterogeneous graph. To consider edge weights in attention mechanism, we involve edge weights while computing attention weights. Other strategies remain the same as the original attention mechanism. The new attention mechanism generates attention weights from the concatenation of $h_a$, $h_b$, where $h_a$ and $h_b$ represent the set of features for node $a$ and $b$, while considering edge weights, in a similar manner with the original attention mechanism. Given $E$ as a set of edges, then every element in $E$ can be represented by a $e_{ab}$ as edge features, where $a$ and $b$ denote the nodes on each end of an edge. $e_{ab}$s are calculated using $W_e$ which represents edge weights between user to item $W_{{e}_{ui}}$ or item to time span $W_{{e}_{is}}$, and any other edge feature in the dataset.

In our model, the attention mechanism $d$ is a single-layer feed-forward neural network, parametrized by a learnable weight vector. 
A Leaky ReLu activation function is then applied. Final coefficients can be expressed as: 

\small
\begin{equation}
    \label{final_att}
    \alpha_{a b}=\frac{\exp \left(\text { LeakyReLU }\left(\overrightarrow{\mathbf{d}}^{T}\left[\mathbf{W} \vec{h}_{a} \| \mathbf{W} \vec{h}_{b}\|\mathbf{W_e} \vec{e}_{ab}\right]\right)\right)}{\sum_{k \in \mathcal{N}_{a}} \exp \left(\text { LeakyReLU }\left(\overrightarrow{\mathbf{d}}^{T}\left[\mathbf{W} \vec{h}_{a} \| \mathbf{W} \vec{h}_{k}\|\mathbf{W_e} \vec{e}_{ak}\right]\right)\right)}
\end{equation}
\normalsize

After calculating attention coefficients, the output of the attention layer is derived by computing the linear combination of node features and corresponding attention coefficients. Non-linearity is then applied:

\small
\begin{equation}
    \label{att_output}
    \vec{h}_{a}^{\prime}=\sigma\left(\sum_{b \in \mathcal{N}_{a}} \alpha_{a b} \mathbf{W} \vec{h}_{b}\right)
\end{equation}
\normalsize
It must be noted that $N_a$ is the set including the first-order neighbors of node $a$ as well as the node $a$ itself.

\subsection{Predictions and Final Aggregation} \label{prediction}

In research, our model is trained for rating prediction. User, item, and time span aggregated information is fed to a feed-forward network to make predictions. Assuming $h^\prime_u$ is the aggregated user features, $h^\prime_i$ is the aggregated item features, and $h^\prime_s$ is the aggregated time span features, the ratings from short-term preference and from long-term preference are predicted using the formula below:

  \begin{equation}
  \noindent
    \label{rating_pred}
    r_l = LeakyReLu\left(W_l\left(h^\prime_u \parallel h^\prime_i\right)\right)
  \end{equation}
  \begin{equation}
    \label{rating_pred}
    r_s = LeakyReLu\left(W_s\left(h^\prime_s \parallel h^\prime_i\right)\right)
  \end{equation}



As a final step, the short-term and long-term predictions are aggregated to complete the model. To form the final prediction, the predictions corresponding to short-term and long-term preferences are aggregated using a linear equation parametrized by learnable weights $\alpha$ and $\beta$ as below:

\begin{equation}
    \label{agg_short_long}
    Y = \alpha r_s + \beta r_l
\end{equation}

$r_s$ and $r_l$ represent the short-term prediction and long-term prediction, respectively. A single-layer neural network is employed to optimize the parameters of $\alpha$ and $\beta$.

\section{Experimental Results} \label{experiments}

\subsection{Datasets and Evaluation Metrics}

We evaluated our model on two real-world datasets: Ciao\footnote{https://www.librec.net/datasets.html} and Epinions\footnote{http://www.trustlet.org/downloaded\_epinions.html}.
Ciao and Epinions \cite{epinion1,epinion2} are datasets 
that contain item ratings given by the users and timestamps when ratings are given. Social relations between the users are also included in the datasets. Users can rate items in the social networks using 1 to 5. Users can add any user to their trust network which creates edges between users. The details of the datasets are given in Table \ref{tab:dataset_attr}. 
In our study, the observation time for both datasets is as the number of seconds from 1/1/1970 0-0-0 to the time the ratings are created.  The item-item edges are created if both items are in the same category. 

\begin{table}[h]
\centering
\caption{Datasets}
\label{tab:dataset_attr}
\begin{tabular}{l|llll}
\hline Dataset &  \#items & \#ratings & \#social & \#users \\\hline
Ciao & 16,121  & 62,452 & 40,133 & 17,589 \\
Epinions & 22,173 & 138,207 & 487,183 & 49,289 \\
\hline
\end{tabular}
\end{table}

Mean Absolute Error (MAE) and Root Mean Square Error (RMSE) are used in this research to evaluate our model and argue its superiority among baselines.



\begin{table*}[h!]\caption{Mean Absolute Error of the proposed model compared to baselines}\label{tab:performance_mae}
\centering\resizebox{\linewidth}{!}{%
\setlength{\tabcolsep}{0.3mm}{
\begin{tabular}{|c|c|c|c|c|c|c|c|c|c|c|c|c|c|c|}

\hline Dataset  & \multicolumn{12}{|c|} { Algorithms } \\
\cline { 2 - 13 } & PMF & SoReg & SoRec & SocialMF & NeuMF  & DeepSoR & GCMC+SN & ScAN & TrustMF & GraphRec & CapsRec & \textbf{Our Model} \\
\hline \hline Ciao(60\%)  & 0.9564 & 0.8973 & 0.8491 & 0.8395 & 0.8295 &  0.7893 & 0.7738 & 0.7631 & 0.7624  & 0.7568 & 0.7585 & \textbf{0.7554} \\
\hline Ciao(80\%) & 0.8972 & 0.8516 & 0.8542  & 0.8303 & 0.7995  & 0.7683 & 0.7507 & 0.7503 & 0.7626 & 0.7409 & 0.7322 & \textbf{0.7302}\\
\hline Epinions(60\%) & 1.0215 & 0.9417 & 0.9091 & 0.8967 & 0.9098 & 0.8517 & 0.8606 & 0.8551 & 0.8546 & 0.8478 & 0.8461 & \textbf{0.8451} \\
\hline Epinions(80\%) & 0.9921 & 0.9172 & 0.8919 & 0.8798 & 0.9023  & 0.8232 & 0.8611 & 0.8443 & 0.8376 & 0.8183 & 0.8149 & \textbf{0.8143} \\
\hline
\end{tabular}}}
\end{table*}

\begin{table*}[h!]\caption{Root Mean Square Error of the proposed model compared to baselines}\label{tab:performance_rmse}
\centering\resizebox{\linewidth}{!}{%
\setlength{\tabcolsep}{0.3mm}{
\begin{tabular}{|c|c|c|c|c|c|c|c|c|c|c|c|c|c|}

\hline { Dataset } & \multicolumn{12}{|c|} { Algorithms } \\
\cline { 2 - 13 } & PMF & SoReg & SoRec & SocialMF & NeuMF  & DeepSoR & GCMC+SN & ScAN & TrustMF & GraphRec & CapsRec & \textbf{Our Model} \\
\hline \hline Ciao(60\%)  & 1.1952 & 1.1034 & 1.0721 & 1.0635 & 1.0849  & 1.0491 & 1.0213 & 1.0186 & 1.0515 & 1.0097 & 1.0072 & \textbf{1.0031} \\
\hline Ciao(80\%) & 1.1194 & 1.0809 & 1.0705 & 1.0574 & 1.0657 & 1.0345 & 1.0249 & 1.0209 & 1.0423 & 0.9815 & 0.9778 & \textbf{0.9762} \\

\hline Epinions(60\%) & 1.2816 & 1.1878 & 1.1498 & 1.1377 & 1.1631 & 1.1208 & 1.0979 & 1.0971 & 1.1524 & 1.0908 & 1.0923 & \textbf{1.0814} \\
\hline Epinions(80\%) & 1.2158 & 1.1635 & 1.1583 & 1.1478 & 1.1497 &  1.0956 & 1.0695 & 1.0686 & 1.1482& 1.0649 & 1.0645 &\textbf{1.0593} \\

\hline

\end{tabular}}}
\end{table*}

\begin{figure*}
\centering
\begin{minipage}{.30\textwidth}
    \centering
    \fbox{\includegraphics[width=0.95\textwidth]{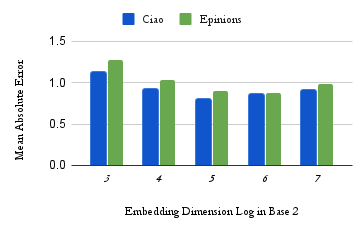}}
    \caption{Embedding Analysis}
    \label{fig:emb_dim_drop_1}
\end{minipage}
\begin{minipage}{.30\textwidth}
    \centering
    \fbox{\includegraphics[width=0.95\textwidth]{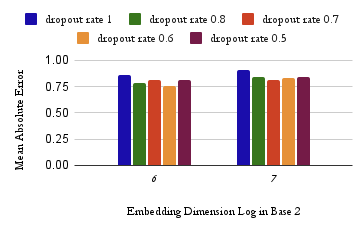}}
    \caption{Dropout Analysis for Ciao}
    \label{fig:dropout_analysis1}
\end{minipage}
\begin{minipage}{.30\textwidth}
    \centering
    \fbox{\includegraphics[width=0.95\textwidth]{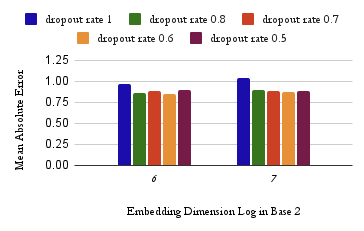}}
    \caption{Dropout Analysis for Epinions}
    \label{fig:dropout_analysis2}
\end{minipage}
\end{figure*}
\begin{figure*}
\centering
\begin{minipage}{.30\textwidth}
    \centering
    \fbox{\includegraphics[width=0.95\textwidth]{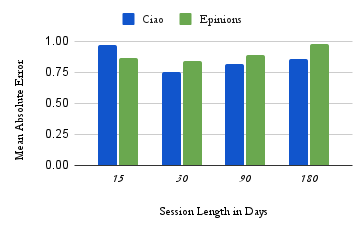}}
    \caption{Time Span Length Analysis}
    \label{fig:session_length}
\end{minipage}
\begin{minipage}{.30\textwidth}
    \centering
    \fbox{\includegraphics[width=0.95\textwidth]{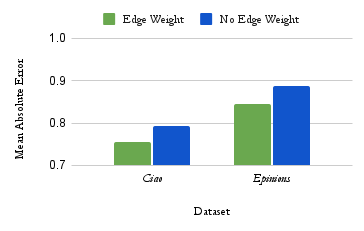}}
    \caption{Edge Weights Analysis}
    \label{fig:edge_weights}
\end{minipage}
\begin{minipage}{.30\textwidth}
    \centering
    \fbox{\includegraphics[width=0.95\textwidth]{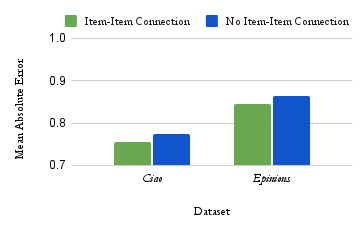}}
    \caption{Item-Item Edge Analysis}
    \label{fig:item-item connection}
\end{minipage}
\end{figure*}
\begin{figure*}[!t]
\centering
\begin{minipage}{.30\textwidth}
    \centering
\fbox{\includegraphics[width=0.95\textwidth]{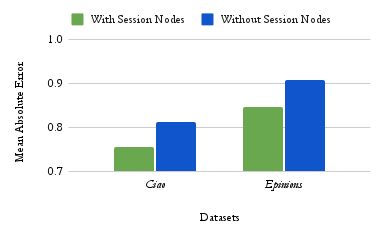}}
    \caption{Time Span Nodes Analysis}
    \label{fig:session_nodes}
\end{minipage}
\begin{minipage}{.30\textwidth}
    \centering
\fbox{\includegraphics[width=0.95\textwidth]{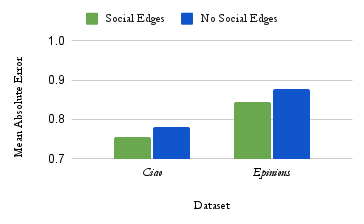}}
    \caption{User-User Edges Analysis}
     \label{fig:social-edges}
\end{minipage}
\end{figure*}

\subsection{Baselines}
We compared our model to eleven state-of-the-art models for recommender systems: PMF \cite{pmf}, SoRec\cite{sorec}, SocialMF \cite{socialmf}, SoReg \cite{soreg}, NeuMF \cite{neumf}, TrustMF \cite{trustmf}, GCMC+SN \cite{gcmc},  DeepSoR \cite{deepsor}, GraphRec \cite{graphrec}, ScAN \cite{scan},   and CapsRec \cite{capsrec}. These baselines adopt different approaches to make rating predictions. Some are neural network-based social recommenders while some others use collaborative filtering methods. There are also some baselines that adopt both methods to enhance the performance of the model. 

\subsection{Hyperparameter Setting}

Both datasets are split into training, validation, and test sets as 60\%, 20\%, 20\%, and 80\%, 10\%, 10\%. To form the GNN model, Leaky ReLu was chosen as the activation function. The default value for embedding size is set to 64 with a dropout rate of 0.6. The default time span length is set to 30 days. The sensitivity of our model to these parameters is explained in the model analysis section. Adam optimization algorithm is used to update network weights iterative based on training data. The train epochs is set to 15. Based on the performances, the short-term $r_s$ and long-term ratio $r_l$ are calculated as 1.154 and 1.217 respectively. 

\subsection{Rating Prediction Performance}

The comparisons between the proposed model and the baselines are mentioned in Tables \ref{tab:performance_mae} and \ref{tab:performance_rmse}. Our model outperforms the compared baselines even with less training data (60\%) on both datasets. The experimental results are repeated for 10 random seed initialization and using T-test, they are statistically significant, with a p-value of less than 0.05. 
It can be found that the lowest accuracy is gained by applying the PMF model. NeuMF, which uses the neural network in collaborative filtering, outperforms PMF. SoRec, SoReg, SocialMF, and TrustMF also surpass the PMF since they apply matrix factorization to both the ratings and social information. The performance of GCMC+SN among other neural network-based baselines including DeepSoR and NeuMF can verify the efficiency of GNNs in graph representation learning. ScAN outperforms all models mentioned above due to the attention mechanism. GraphRec performed better than ScAN because of the additional item space aggregations while employing the attention mechanism. CapsRec performs slightly better than GraphRec and outperforms all other baselines which implies the efficiency of using capsule networks.

\subsection{Model Analysis} \label{Ablation}

We investigated the performance of our model by removing or changing certain components and changing the parameter settings to understand the contribution of the model settings. The ablation study is applied to the Ciao (60\%) and Epinions (60\%) datasets.

\subsubsection{Embedding Dimensions and Dropout Rate} 
The embedding size can have a great effect on the performance. The sensitivity of the proposed model to embedding sizes is measured and compared. 
As we reduce the embedding size, the performance decreases since the model cannot represent all necessary features in the embeddings. Thus, the model would not gain rich information from the nodes to make predictions. However, increasing the number of embedding dimensions to more than a threshold will increase the risk of over-fitting, and generalization problem will arise. To solve that, we employed the dropout regularization method while setting large numbers as the embedding size.

We set the dropout rate to 1 and change the embedding sizes from 8 to 128 to observe the impact. 
As shown in Figure \ref{fig:emb_dim_drop_1}, without using dropout regularization, embedding size 32 for Ciao and 64 for Epinions dataset produce less error. Then, for large sizes 64 and 128, we applied different dropout rates to evaluate the performance. 
The results illustrated in Figures \ref{fig:dropout_analysis1} and \ref{fig:dropout_analysis2} show that using embedding size 64 with a dropout rate of 0.6 produces less error on both data sets. This study demonstrates the importance of tuning the dropout rate with the embedding dimensions together to obtain optimum results.

\subsubsection{Length of Time Span} The time span node is the main element to define short-term preferences. In this method, the length of the time span needs to be defined. To find the best value, we experimented different time span lengths including 15, 30, 90, and 180 days. All the other parameters are set to default except the one understudy.

Figure \ref{fig:session_length} shows the results. For both datasets, when the time span length is set to 30 days,  the returned MAE is the lowest. With the increase in the length of time span, the preferences of the user can change dramatically. Thus, the time span node cannot capture the short-term preference very well. Also, reducing the time span beyond some point could result in too many time span nodes which makes the graph computationally heavy. Moreover, many users do not make any interactions during very short periods.

\subsubsection{Edge Weights} In this research, we proposed adding the edge weights to the graph to capture the importance of interactions occurring closer to the prediction time. We evaluated the sensitivity of our model to the edge weights by removing them from the graph and remeasuring the performance metrics. The result is illustrated in Figure \ref{fig:edge_weights}. It shows that edge weights clearly have a significant contribution to improved performance.  Without adding the edge weights, the MAE is much higher. 

\subsubsection{Item-Item Connection} We evaluated the effect of the item-item connections by measuring the performance of the model without them. The results illustrated in Figure \ref{fig:item-item connection} argue that although adding item-item connections makes the graph more complex, it is worth considering since it enhances performance by bringing more information to the graph.

\subsubsection{Time Spane (Session) Nodes} To prove that adding time span nodes can improve the performance, we measured the performance of the model without time span nodes. The results are compared to the model performance with time span nodes. The comparison is illustrated in Figure \ref{fig:session_nodes}. It shows that adding time span nodes was effective in reducing errors.

\subsubsection{User-User Connection} The contribution of social edges in the system is also evaluated. The comparison illustrated in Figure \ref{fig:social-edges}. The results show that adding the user-user (social) edges to the model can improve the performance significantly.

\section{Conclusion}\label{conclusion}

User preferences change over time. Capturing short-term and long-term preferences in a graph poses great challenges. In this research, we proposed a novel model to make recommendations through modeling long-term and short-term preferences by adding time span nodes to a graph. Our model employs the attention mechanism on a new graph structure containing time span nodes and edge weights in addition to the basic components of a traditional social graph. Experiments confirm the effectiveness of the proposed approach over other existing baselines. Future work includes utilizing time span nodes in a dynamic graph structure and personalizing the time span nodes to each user to better capture user preferences.

\bibliographystyle{IEEEtran}
\bibliography{conference_101719}

\end{document}